\documentclass[aps,prl,showpacs,floatfix,notitlepage,a4paper,twocolumn]{revtex4}

\usepackage{amsfonts,amsmath,units,wasysym,epsfig,graphicx,verbatim,color,subfigure,graphicx,bm,mathrsfs,lipsum,hyperref}
\usepackage[utf8]{inputenc}
\usepackage[normalem]{ulem}  

\begin{document}

\newcommand{\UNIPI}{
Dipartimento di Fisica ``Enrico Fermi'', Universit\`a di Pisa, Largo B. Pontecorvo 3, I-56127 Pisa, Italy
}

\newcommand{\INFNPI}{INFN Sezione di Pisa, Largo B. Pontecorvo 3, I-56127 Pisa, Italy}

\newcommand{\UNIFE}{Dipartimento di Fisica e Scienze della Terra,  Università di Ferrara, Via Saragat 1, I-44122 Ferrara, Italy}
\newcommand{\INFNFE}{INFN Sezione di Ferrara, Via Saragat 1, I-44122 Ferrara, Italy}

\newcommand{\INFNCT}{
INFN Sezione di Catania,  Via Santa Sofia 64, I-95123 Catania, Italy
}

\title{Was GW190814 a black hole -- strange quark star system?}

\author{I. Bombaci$^{1,2}$, A. Drago$^{3,4} $, D. Logoteta$^{1,2}$, G. Pagliara$^{3,4}$ and I. Vida\~na$^5$ }

\affiliation{$^1$\UNIPI,$^2$\INFNPI,$^3$\UNIFE,$^4$\INFNFE,$^5$\INFNCT}


\begin{abstract} 
We investigate the possibility that the low mass companion of the black hole in the source of GW190814 was a strange quark star. This possibility is viable within the so-called two-families scenario in which neutron stars and strange quark stars coexist. Strange quark stars can reach the mass range indicated by GW190814, $M\sim (2.5-2.67) M_\odot$ due to a large value of the adiabatic index, without the need for a velocity of sound close to the causal limit. Neutron stars (actually hyperonic stars in the two-families scenario)
can instead fulfill the presently available astrophysical and nuclear physics constraints which require a softer equation of state. 
In this scheme it is possible to satisfy both the request of very large stellar masses and of small radii while using totally realistic and physically motivated equations of state. Moreover it is possible to get a radius for a 1.4 $M_\odot$ star of the order or less than 11 km, which is impossible if only one family of compact stars exists.

\end{abstract}
\maketitle

\section{Introduction}
The gravitational wave signal GW190814 \cite{Abbott:2020khf} 
has been generated by the merger of a binary system whose components are a $23\,M_{\odot}$ black hole (BH) and a  $(2.5-2.67)\,M_{\odot}$ compact object.   
Explaining the nature of this compact object is nowadays a big challenge for astrophysics and dense matter physics. The value of its mass falls within the expected lower end of the so-called mass gap ($2.5\,M_{\odot} < M < 5\,M_{\odot}$) and, therefore, this object is not expected to be a BH. Nevertheless, there are interpretations based on the assumption that GW190814 is actually a binary BH system, and proposals for the formation of light BHs (including primordial BHs) have been put forward \cite{Safarzadeh:2020ntc,Yang:2020xyi,Clesse:2020ghq,Vattis:2020iuz}.
In contrast, if this compact object is instead a neutron star (NS), then several issues concerning the stiffness of the equation of state (EOS) and the rotational properties of NSs need to be addressed \cite{Tsokaros:2020hli}. 

There are currently two possible explanations for the existence of such a massive NS: the EOS of dense matter is significantly stiff in order to support such large mass \cite{Fattoyev:2020cws,Tews:2020ylw,Lim:2020zvx,Sedrakian:2020kbi,Huang:2020cab} or the NS was rotating very close to the keplerian limit \cite{Most:2020bba,Zhang:2020zsc,Dexheimer:2020rlp}. There are, however, two major drawbacks of these kinds of interpretations. In the first case one needs a rather stiff EOS, 
with a speed of sound $v_s$ that should exceed $\sqrt{0.6}c$ \cite{Tews:2020ylw}. 
In turn, EOSs which allow for the existence of such massive NSs are in tension with constraints obtained from heavy-ion collisions experiments \cite{Danielewicz:2002pu}  and from the tidal deformability constraints derived from GW170817 \cite{Fattoyev:2020cws} which favor softer EOSs.
Regarding the second explanation, one needs to explain how a BH-NS system could merge before dissipating the large natal NS angular momentum \cite{Abbott:2020khf}.
Other possibilities are based on extended gravity theories \cite{Astashenok:2020qds,Moffat:2020jic} 
or on theories predicting the existence of bosonic stars or gravastars, see \cite{Abbott:2020khf} and references therein.

We propose a solution that does not require 
new physics ({\it e.g.,} modified gravity or the inclusion of new particles) apart from the assumption that the true ground state of strongly interacting matter is not $^{56}$Fe but a deconfined mixture of up ($u$), down ($d$) and strange ($s$) quarks, namely strange quark matter (QM)  \cite{bodmer71,witten84,terezawa89a,terezawa89b}.
We investigate the possibility that indeed the low mass component in GW190814 was not a BH nor an {\it ordinary} NS 
but a (strange) quark star (QS), {\it i.e.,} a star entirely composed of deconfined $u,d,s$ QM.  
Calculations considering only lowest-order perturbative interactions between quarks produce very compact  
($R_{1.4}\lesssim 11\,$km, where $R_{1.4}$ is the radius of a star having 
a gravitational mass $M = 1.4\,M_\odot$) and not too massive ($M_{max}^Q < 2\,M_{\odot}$) QSs \cite{Lattimer:2000nx}. 
The introduction of more sophisticated quark dynamics has indicated the possibility that the mass of QSs could reach larger values,  $M_{max}^Q\sim 2.75 \,M_{\odot}$ \cite{Kurkela:2009gj}, what became extremely relevant after the discovery of compact stars with $M \sim 2\,M_{\odot}$. Also, those large masses can be reached without the need for sound velocities close to the causal limit because in QSs the adiabatic index diverges at the surface of the star \cite{Haensel:1986qb}.
It is commonly accepted however that not all compact stars can be QSs, for instance magnetar oscillations pose challenges for QSs \cite{Watts:2006hk}. 

NSs and QSs could coexist,  
as has been proposed and discussed in detail in several papers \cite{bombaci_datta_2000,Berezhiani:2002ks,Bombaci:2004mt,Drago:2004vu,Drago:2013fsa,Drago:2015dea,Drago:2015cea,Bombaci:2016xuj,Drago:2020gqn}.  
This is a viable possibility for relieving the possible tension between the indications of the existence of very compact stars, $R_{1.4}\lesssim 11.5\,$km,  
suggested by some analyses on thermonuclear bursts and X-ray binaries
and the existence of very massive stars \cite{Drago:2013fsa,Drago:2020gqn}.
Actually, if the low mass component of GW190814 is a NS, it has been shown that, if only one family of compact stars exists, $R_{1.4} \gtrsim (11.6-11.8)\,$km due to the causal limit \cite{Lim:2020zvx,Godzieba:2020tjn}. Moreover, those rather small radii are obtained only in 
the extreme situation in which most of the star is occupied by matter with $v_s \sim c$. 
Two coexisting families of compact stars are thus necessary if 
$M_{max} \sim 2.6\, M_\odot$ and $R_{1.4} \lesssim 11.6\,$km.

\section{Equations of State}
Let us discuss the EOSs of hadronic matter (HM) and of QM that could possibly describe the two coexisting families of compact stars.
 
The first EOS of HM (composed of nucleons and hyperons)  
is obtained within the Brueckner-Hartree-Fock (BHF) approach using nucleon-nucleon 
and three-nucleon interactions derived in chiral effective field theory suplemented by 
nucleon-hyperon and nucleon-nucleon-hyperon interactions \cite{Logoteta:2019utx}.  
This microscopic EOS reproduces the empirical properties of nuclear matter at the saturation density 
$n_0 = 0.16\,\mathrm{fm}^{-3}$, does not violate causality ({\it i.e.,} $v_s < c$), and is consistent (see Fig.\ 2 in \cite{BL2018}) with the measured elliptic flow of matter in 
heavy-ion collisions experiments \cite{Danielewicz:2002pu}.   
When computing the mass-radius (M-R) relation for the corresponding {\it ordinary} NSs (also 
referred to as hadronic stars (HSs), {\it i.e.} compact stars with no fraction of QM) we obtain:  
i) a maximum mass $M^H_{max} \sim 2\,M_{\odot}$ (the transition to a QS is discussed later);
ii) a tidal deformability of the $1.4\,M_\odot$ configuration $\Lambda_{1.4} = 388$, compatible with 
the constraints derived from GW170817 \cite{TheLIGOScientific:2017qsa}; and
iii) a threshold mass, for the prompt collapse to a BH of the postmerger compact object in GW170817,  $M_{\mathrm{threshold}} = 2.79\, M_{\odot}$ 
(estimated by using the empirical formula given in Ref.\ \cite{Koppel:2019pys}) indicating that GW170817 is compatible with being a NS-NS system if NSs are described by this EOS.

The second EOS we consider is based on a relativistic mean field (RMF) scheme in which nucleons, hyperons and $\Delta$-resonances are present \cite{Drago:2014oja}. The effect of the production of $\Delta$s is a further softening of the EOS: from one side this allows values of $R_{1.4}$ as small as $11\,$km \cite{Drago:2013fsa,Burgio:2018yix} to be reached, but on the other side the maximum mass is limited to be $M^H_{max} \sim 1.6\,M_{\odot}$.  
The tidal deformability is $\Lambda_{1.4}= 150$ and $M_{\mathrm{threshold}}\sim 2.5\, M_{\odot}$ \cite{DePietri:2019khb}. Therefore, when using this EOS, GW170817 cannot be described as a NS-NS merger, but it can be a NS-QS merger. In that case the average tidal deformability associated with the mixed binary system is $\tilde \Lambda \sim 450-550$, depending on the adopted quark EOS \cite{Burgio:2018yix}, and $M_{\mathrm{threshold}}\sim (3- 3.5)\, M_{\odot}$, again depending on the quark EOS. 

\begin{figure}[t]
	\begin{centering}
		\epsfig{file=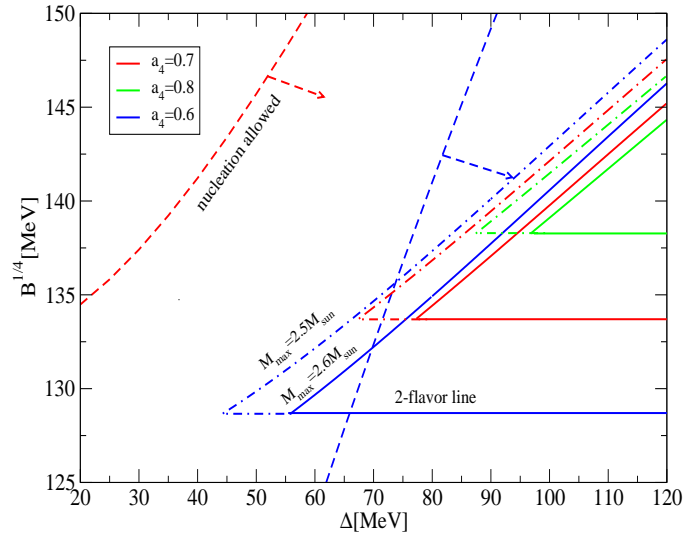,height=8.8cm,width=7cm,angle=-90}
		\caption{
The triangular regions delimit the values of $\Delta$ and $B^{1/4}$ which lead to 
$M_{max}^{Q} \geq 2.6\,M_\odot$ (solid lines)  and $M_{max}^{Q} \geq 2.5\,M_\odot$ (dot-dashed lines) for three different values of $a_4$. 
The horizontal lines indicate the values of $B^{1/4}$ below which two-flavor QM is more stable than $^{56}$Fe and are thus excluded, see analogous figure in \cite{Weissenborn:2011qu}. The dashed lines indicate the regions of the parameter space for which QM nucleation is possible for the 
$1.5\, M_{\odot}$ metastable HS configuration obtained with the RMF EOS. For the BHF EOS those lines are not shown because they do not further restrict the parameter space with respect to the constraint on $M_{max}^Q$. 
}
			
	\label{f:suppl1}
	\end{centering}
\end{figure}

For the second family of compact stars, QSs, we use the simple QM model suggested 
in Ref.\  \cite{Alford:2004pf} where the grand canonical potential reads \footnote{This potential can be obtained by expanding the potential proposed in \cite{Alford:2002rj} in terms of the ratio between the strange quark mass and the quark chemical potential.}: 
\begin{equation}
\Omega=-\frac{3}{4 \pi^2}a_4\mu^4+\frac{3}{4 \pi^2}\mu^2(m_s^2-4\Delta^2)+B
\label{eosquark}
\end{equation}
 being $\mu$ the quark chemical potential, $B$ the bag constant, $a_4$ a parameter that encodes perturbative QCD corrections, $\Delta$ the color–flavor locking (CFL) superconducting gap \cite{Alford:2007xm} and $m_s$ the strange quark mass.  The parameter space allowing for $M_{max}^{Q} \geq 2.6\,M_\odot$ ($\geq 2.5\,M_\odot$, solid and dot-dashed lines respectively) is displayed in Fig.\ \ref{f:suppl1}. In the following discussion, as an example, we consider the parameter set $B^{1/4} = 135\,\mathrm{MeV}$, $a_4 = 0.7$, $\Delta = 80\,\mathrm{MeV}$ and $m_s = 100\,\mathrm{MeV}$ that allows $M_{max}^{Q} = 2.58\, M_{\odot}$.
Notice that the sign of $a_2=m_s^2-4\Delta^2$ for this choice of parameters is negative. Negative values of $a_2$ have been explored also in Ref.\cite{Alford:2002rj} where $\Delta$ is considered to be as large as $150$ MeV and $m_s$ is varied in the range $(150-300)$ MeV. A negative value of $a_2$ reduces the effective bag constant which in turn favors the existence of quark stars and disfavors the formation of hybrid stars, as found in Ref.\cite{Alford:2002rj}. 

\begin{figure}[!ht]
	\begin{centering}
		\epsfig{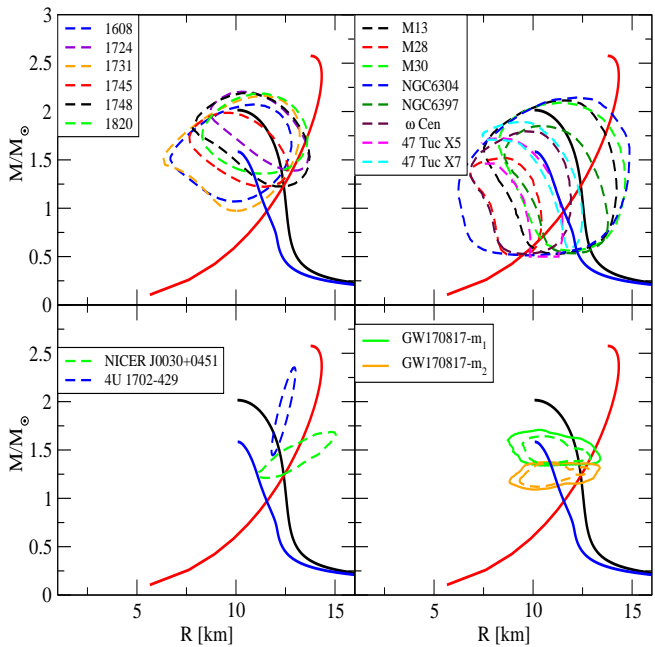}
		\caption{Mass-radius relations of NSs (thick black and blue lines correspond to the 
		BHF and RMF EOS, respectively) and of QSs (thick red line), compared with several astrophysical data. 
		Upper left panel: thermonuclear bursts \cite{Ozel:2016oaf}. Upper right panel: low mass X-ray binaries in quiescence \cite{Ozel:2016oaf}. Lower left panel: the NICER results for J0030+0451 \cite{Miller:2019cac} and the RXTE results for the cooling tail spectra of 4U1702-429 \cite{Nattila:2017wtj}. We have modelled the mass-radius posterior distribution of 4U1702-429 by using a bivariate gaussian with $\rho=0.9$ as in \cite{Jiang:2019rcw}.
		All constraints are at the $68\%$ CI.
		Lower right panel: constraints on the two compact stars of GW170817, solid (dashed) lines correspond to the $90\%$ ($68\%$) CI \cite{Abbott:2018exr}.}
			
	\label{f:fig1}

	\end{centering}
\end{figure}


It is well known that in the limit of massless, non-interacting quarks the maximum mass of QSs 
scales as $B^{-1/2}$ \cite{witten84} and, therefore, to obtain very massive QSs 
one has to adopt small values of the bag constant. In turn, such small values of $B$ easily lead to an unreasonably small critical density for the phase transition to QM, which is excluded by 
heavy-ion experiments \cite{Danielewicz:2002pu}. By using 
Eq.\ (\ref{eosquark}) we avoid that problem, because the CFL gap can exist only if strange quarks are present and abundant \cite{Lugones:2002va,Bombaci:2006cs}.

We explicitly checked that at $T = 0$ nuclear matter is energetically favored with respect 
to two-flavor QM up to the highest density reached in the computed EoS $\sim 7 n_0$.

\vskip 1cm
\begin{figure}[!ht]
	\begin{centering}
		\epsfig{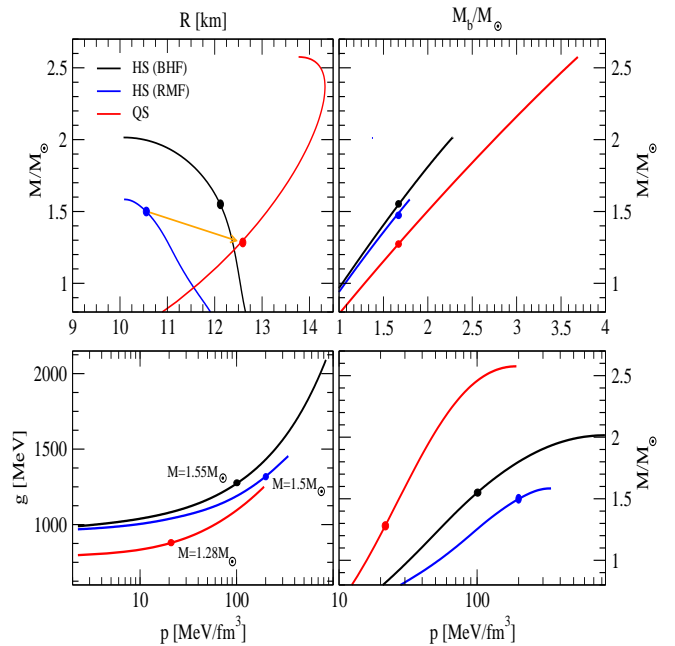}

		\caption{Upper left panel: gravitational mass vs radius for HSs (black and blue lines) and QSs (red line).
		Also shown an example of jump (orange arrow) from the first to the second family of stars.  
		Upper right panel: gravitational mass $M$ vs baryonic mass $M_b$ for HSs and QSs.  
		QSs are more bound than HSs with the same baryonic mass. 
		Lower left panel: Gibbs energy per baryon $g_H$ ($g_Q)$ for the hadronic (quark) phase  
		vs pressure. Since, at fixed pressure, $g_Q < g_H$ then  QM nucleation is allowed. 
		Lower right panel: gravitational mass vs central pressure for HSs and QSs. 
		The dots in all panels indicate the critical hadronic configuration, for each EOS model, and the final QS configuration having the same  baryonic mass of the initial one.   
		Notice that the final QS configurations are very close in mass (see text) and thus 
		both are indicated by the same red dot. }

	\label{f:fig2}
	\end{centering}
\end{figure}

\section{Mass-radius relations}
We display in Fig.\ \ref{f:fig1} the M-R relations for HSs and  QSs, compared with several astrophysical constraints. The two hadronic EOSs are soft enough to satisfy most of the constraints suggesting small radii for stars having a mass of about $(1.4-1.5 )M_\odot$. The RMF EOS can reach particularly small radii, displaying a $R_{1.4}< 11$ km, but it is not clear if this is really needed because the constraints on masses and radii depend strongly on the composition of the stellar atmosphere, and larger radii are obtained when assuming a He rather than a H atmosphere, see {\it e.g.,} \cite{dEtivaux:2019cnf}. It is clear that M-R relations giving values of $R_{1.4}$ in the range (11-12) km can be obtained with EOSs sitting in between the BHF and the RMF. 
Notice that in the case of the RMF EOS the corresponding $M_{max}^H$ is smaller than $2 M_\odot$.  
This is not a problem within the two-families scenario 
since $2\, M_\odot$ compact stars belong to the QSs family.  
It is interesting to note that the constraints coming from NICER and from GW170817 can be satisfied by using both BHF and RMF EOSs, but as already mentioned, in the case of BHF GW170817 was a NS-NS merger (although a NS-QS merger could also be possible) whilst when using the RMF EOS GW170817 can only be explained as an NS-QS merger  \cite{Drago:2017bnf,DePietri:2019khb,Burgio:2018yix}. The constraint from NICER suggests either an HS or a QS when using the BHF EOS whilst in the case of the RMF EOS the QS interpretation seems more likely. Finally, 4U 1702-429 limits are satisfied by the BHF EOS, whereas the interpretation as a QS seems more problematic, at least with the present quark EOS.

\section{Conversion of a hadronic star into a quark star}  
It has been shown \cite{Berezhiani:2002ks,Bombaci:2004mt,Drago:2004vu} that HSs above a threshold 
value of their mass (threshold central density $n_{thr}$) become metastable to the conversion to QSs. Metastable HSs have a {\it mean-life time} related to the nucleation time $\tau$ to form the first critical-size QM droplet in their center. 
As shown in \cite{Berezhiani:2002ks,Bombaci:2004mt,Drago:2004vu}, $\tau$ decreases very steeply 
with the stellar central density $n_c$  
from $\tau = \infty$ (when $n_c = n_{thr}$) to values much smaller than typical pulsar ages 
(Fig. 1 in \cite{Berezhiani:2002ks}).  
At this point (e.g. when $\tau = 1 \,$yr) the conversion to a QS is very likely. 
This configuration defines \cite{Berezhiani:2002ks} the {\it critical mass}   
$M_{cr} = M^H(n_{cr}) = M^H(\tau = 1 \,yr)$ and the corresponding critical density $n_{cr}$. 
This conversion process releases a huge amount of energy of $\sim 10^{53}\,\mathrm{erg}$\, \cite{bombaci_datta_2000}. 
A way to produce QSs is therefore through mass accretion onto HSs in binary systems   \cite{Berezhiani:2002ks,Wiktorowicz:2017swq} or during the spin-down of rapidly rotating HSs    
\cite {Bhattacharyya:2017mdh}.  
Within the two-families scenario QSs can also be produced through the merger of two HSs  \cite{Drago:2017bnf,DePietri:2019khb} and, potentially, through the conversion of a 
protoneutron star (PNS) into a QS immediately after a successful core-collapse
supernova \cite{Fischer:2010wp} or during the early evolution of a PNS \cite{Bombaci:2011mx,Lugones:2015bya}. 
The exact value of $M_{cr}$ and $n_{cr}$ needs a 
calculation  
of the nucleation time of the first QM droplet with the same flavor content 
as that of the HM in the stellar center  \cite{Berezhiani:2002ks,Bombaci:2004mt,Drago:2004vu,Bombaci:2016xuj,Bombaci:2008wg}  
and is temperature-dependent when considering PNSs 
\cite{Bombaci:2011mx,Bombaci:2016xuj}.
Here for simplicity we choose as the critical density the one for which the hyperon fraction  
reaches the value $n_Y/n_B = 0.10$ in the stellar center, 
namely $n_{cr} \sim 0.9\,(0.5)$ fm$^{-3}$ for the RMF (BHF) EOS.  
The corresponding  critical HS configuration is marked by a dot on the HSs curves in all panels of 
Fig.\ \ref{f:fig2}. 
Once the process of deconfinement starts it proceeds at first as a rapid deflagration (notice that the central pressure of the QS is smaller than the pressure of the original HS, lower right panel in 
Fig.\ \ref{f:fig2}) and then as a diffusion \cite{Drago:2015fpa}: during this process the baryon number is conserved (upper right panel), but the gravitational mass decreases because the process is strongly exothermic. 
The critical baryonic masses are of $1.67\,M_{\odot}$ and $1.68\, M_{\odot}$ for the BHF and the RMF EOSs, respectively. Thus, the final QSs configurations are very close in mass and therefore both indicated by the same dot. 
In our scenario the distribution of HSs is thus restricted to masses smaller than about $(1.5-1.6) M_\odot$, while QSs have masses larger than $\sim 1.3 M_\odot$. Interestingly, this partition is very similar to the one suggested by the analysis of the distribution of the masses of NSs \cite{Antoniadis:2016hxz,Tauris:2017omb,Alsing:2017bbc}. Also, the number of QSs in binary systems produced by mass accretion onto an HS is rather limited because they form from HSs close to the critical mass: in \cite{Wiktorowicz:2017swq} the fraction of QSs in LMXBs was estimated to be smaller than about 25\%. The fraction of HS-QS and QS-QS systems is even smaller. 

The possibility for an HS to convert into a QS with larger radius has been analysed in detail in  \cite{Drago:2020gqn}. The total binding energy of compact stars is the sum of the gravitational 
and nuclear binding energies \cite{bombaci_datta_2000}, the last being related to the microphysics 
of the interactions. An HS can convert into a QS with a larger radius because the consequent reduction 
of gravitational binding is overcompensated by the large increase in the nuclear binding and 
the process turns out to be exothermic. 

\section{Discussion and conclusions}
We have shown that within the two-families scenario it is possible to have compact stars reaching a mass similar to the one indicated by GW190814 and that those massive objects are QSs. We can obtain that result while using physically motivated EOSs for the hadrons and for the quarks and without the need to assume velocities of sound close to the causal limit. 
A sizable superconducting gap in the quark phase is needed, still in the range indicated in the literature.
Finally, the hadronic branch can have very small radii, breaking the limits derived by assuming that only one family of compact stars exists.

Which are the possible evolutionary paths leading to the formation of a $(2.5-2.6) M_\odot$ QS in GW190814? One possibility is that GW190814 originates from a triple system \cite{Liu:2020gif,Lu:2020gfh} and the QS formed from the merger of two lighter NSs. Another possibility is that it was produced directly as a heavy QS from an anomalous supernova explosion powered by quark deconfinement, as mentioned above. 

What are the implications of $M_{max}\sim (2.5-2.6) M_\odot$? 
Since the distribution of masses of compact stars in binary systems is peaked around $1.33M_{\odot}$ with $\sigma = 0.11\, M_{\odot}$ \cite{Kiziltan:2013oja} a large fraction of mergers would produce a stable or a supramassive QS. 
In the case of GW170817 $M_{tot}\sim 2.73 M_\odot$, the outcome of that merger was most likely a stable QS and not a BH. Only a small amount of rigid rotation, if any, would be needed to avoid collapse to a BH since the mass of the final object is smaller than $M_{tot}$ due to the ejection of matter and of energy in the form of GWs and neutrinos. This possibility would fit with the suggestion that a long-lived NS was the outcome of GW170817 \cite{Piro:2018bpl} although the observed prolongued X-ray emission could also be related to  the non-thermal “kilonova  afterglow” \cite{Troja:2020pzf}.
Notice that limits on the lifetime of the GW170817's remnant have been put in \cite{Lippuner:2017bfm,Margalit:2017dij} by examining the nucleosynthesis of heavy elements. It has been shown that the remnant should collapse within a few tens of ms to avoid too large a deposition of energy in the ejecta. Notice anyway that if the conversion to QS occurs immediately after the merger, the moment of inertia of the newly formed object increases and a significant fraction of the rotational energy dissipates into heat 
\cite{Pili:2016hqo}. In  this  way the energy transferred to the kilonova is reduced, alleviating  the  problems of a delayed collapse to BH. Moreover, in the two-families scenario GW170817 could be interpreted 
as a HS-QS system \cite{Drago:2017bnf,DePietri:2019khb}, but a detailed investigation on the associated nucleosynthesis is still missing.

While these implications of $M_{max}\sim (2.5-2.6) M_\odot$ are rather general and do not refer specifically to a QS, a more direct implication can be drawn when considering possible mechanisms for the generation of short gamma-ray bursts. Clearly, if the outcome of the merger is a stable or a supramassive NS or QS, mechanisms based on the formation of a BH would not be appropriate, but one can consider the protomagnetar model \cite{Thompson:2004wi,Metzger:2010pp,Metzger:2007cd,Bucciantini:2011kx}. In that model though a problem exists concerning the duration of the burst, which is related to the time during which a jet with the appropriate baryon load is produced: typically that time is of the order of seconds or tens of seconds if the surface of the star is made of nucleons. In Ref.\ \cite{Drago:2015qwa} it has been shown that, if during the merger a QS is produced, then that time reduces to tenths of a second since baryons stop being ablated and ejected once the surface of the star converted into quarks. Therefore our suggestion that the binary system generating GW190814 contained a QS is consistent with a global scenario of gamma-ray burst production.

\bibliography{ref}
\end{document}